\documentclass{optica-article}

\journal{opticajournal} 

\articletype{Research Article}
\usepackage{caption,subcaption}
\usepackage{graphicx}
\usepackage{lineno}
\usepackage{hyperref}
\usepackage{cleveref}
\usepackage[dvipsnames]{xcolor}

\begin{document}

\title{Fast Wide-field Light Sheet Electro-optic FLIM}

\author{V. Rose Knight\authormark{1}, Nils Bode\authormark{2}, Dara P. Dowlatshahi\authormark{3}, Jung-Gun Kim\authormark{4}, Mary Beth Mudgett\authormark{4}, Soichi Wakatsuki\authormark{3}, Adam J. Bowman\authormark{5,*}, and Mark A. Kasevich\authormark{1,*}}

\address{\authormark{1}Physics Department, Stanford University, 382 Via Pueblo Mall, Stanford, California 94305, USA\\
\address{\authormark{2}Department of Physics, Friedrich-Alexander University, Erlangen, Germany\\
\authormark{3}SLAC National Accelerator Laboratory, 2575 Sand Hill Road, Menlo Park, California 94025, USA\\
\authormark{4}Department of Biology, Stanford University, Stanford, California 94305\\
\authormark{5}Salk Institute for Biological Studies, 10010 N Torrey Pines Rd, La Jolla, CA 92037, USA}}

\email{\authormark{*}abowman@salk.edu; kasevich@stanford.edu} 


\begin{abstract*} 
We demonstrate volumetric fluorescence lifetime microscopy (FLIM) using the electro-optic FLIM technique. Images acquired in a selective plane illumination microscope are gated using a Pockels cell driven at 80 MHz, enabling light sheet electro-optic FLIM (LS-EO-FLIM) acquisition with up to 800 $\mu$m field of view. Volume acquisitions are demonstrated on fluorescent bead mixtures and in live \textit{Arabidopsis thaliana} root samples using both genetically encoded fluorescent proteins and endogenous autofluorescence. 
\end{abstract*}

\section{Introduction}

Fluorescence lifetime microscopy (FLIM) enables acquisition of quantitative, label-free, and
chemically specific contrast in a variety of biological imaging contexts \cite{berezin_fluorescence_2010, datta_fluorescence_2020}. Physical and chemical factors that modify the environment of a fluorescent probe molecule modulate the available nonradiative decay pathways and change its fluorescence lifetime. FLIM techniques
are devoted to resolving these changes, and they have traditionally found uses in a variety of
two-dimensional applications including imaging fluorescent biosensors, molecular probes,
and autofluorescent indicators of cellular metabolism. Lifetime imaging is particularly promising
in live imaging studies of plants where strong autofluorescence is present — for example from lignin and chlorophyll — and where it is important
to effectively differentiate diverse endogenous species and fluorescent probe signatures \cite{donaldson_autofluorescence_2020, escamez_fluorescence_2021}. 

While FLIM is a powerful modality, it has remained challenging to combine with three-dimensional imaging methods due to the limited throughput of time-correlated single photon counting (TC-SPC) detectors and the large number of points which must be sampled. Despite this difficulty, recent works have identified the promise of combining FLIM with light sheet microscopy (LS-FLIM) due to light sheet’s ability
to provide volumetric imaging with low phototoxicity. Light sheet fluorescence microscopy has also found powerful application visualizing dynamic processes such as germination of plant seedlings across spatial scales \cite{ovecka_multiscale_2018}. The detectors used in previous LS-FLIM microscopes
still involve fundamental trade-offs in either photon throughput or pixel noise performance. These
include microchannel plates with cross-wire \cite{hirvonen_lightsheet_2020} or spatially-resolved anodes \cite{birch}, modulated
complementary metal-oxide semiconductor (CMOS) cameras \cite{mitchell_functional_2017}, gated optical intensifiers \cite{li_digital_2020, funane_selective_2018, weber_monitoring_2015, Greger2011}, and single-photon avalanche detector (SPAD) arrays \cite{samimi_light-sheet_2023, nutt}. Challenges have
included long acquisition times of ten seconds or even minutes per slice \cite{hirvonen_lightsheet_2020} and imaging
noise that requires a reference intensity image to be acquired on a standard scientific camera \cite{samimi_light-sheet_2023}. Recently developed $512\times512$ SPAD arrays operating in a gated mode rather than TC-SPC have allowed for scanned light sheet acquisition with one second exposures per plane but still face significant detection efficiency and throughput limitations compared to scientific camera sensors, requiring correction for photon pile-up, efforts to avoid detector saturation, and bandwidth limitations \cite{Dunsing-Eichenauer2025-cv}. 

Here we adapted the recently developed electro-optic FLIM (EO-FLIM) technique \cite{bowman_electro-optic_2019, bowman_resonant_2021, bowmanscience} to enable an alternative
approach to light sheet FLIM which is compatible with high photon throughput and dynamic range while maintaining
the favorable pixel noise and image quality of scientific cameras for imaging low-photon yielding fluorescent
samples. The method exploits electro-optic switching of the image plane in combination with pulsed laser excitation to obtain wide-field FLIM images.  The method is straightforward to implement in existing fluorescent microscope systems, and does not require specialized image detectors or dedicated high-speed signal processing electronics. This approach  is compared with existing work in Supplemental Table 1.

\section{Methods}
\begin{figure}[htbp!]
\centering\includegraphics[width=\textwidth]{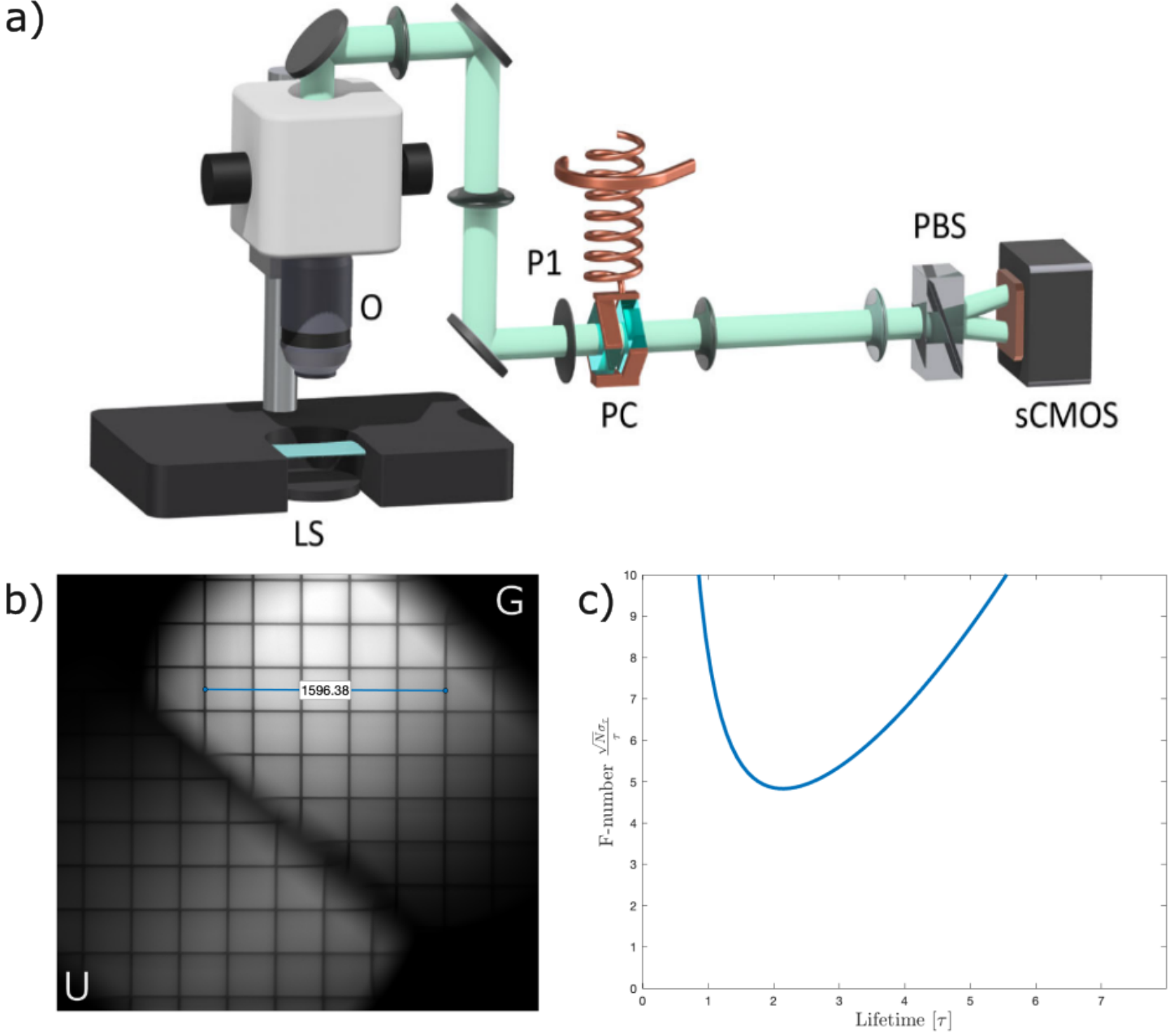}
\caption{Details of the experimental LS-EO-FLIM setup. (a) Schematic showing imaging path through commercial microscope body, resonant inductor-capacitor (LC) circuit for driving Pockels cell, and polarization splitting with a Wollaston prism before the camera. Abbreviations: LS, light sheet; O, objective; P1, polarizer; PC, Pockels cell; PBS, polarizing beam splitter; sCMOS, scientific CMOS camera.
(b) Photon-normalized lifetime estimation sensitivity (as quantified by the F-number, the ratio of the estimated uncertainty to the fundamental limit associated with photon shot noise), a function of lifetime, computed at the maximally sensitive phase delay from experimental IRF. All measurements were taken at this phase delay. (c) Output of 
camera sensor imaging a 200 $\mu$m grid calibration sample shows gated G (s-polarized path from Pockels cell) and ungated U (p-polarized path from Pockels cell) images. Scale bar: 1596.38 pixels, approximately 1 millimeter.}
\label{fig:setup}
\end{figure}

The EO-FLIM technique uses a Pockels cell to implement nanosecond optical gating of fluorescence photons across
a wide-field image. In order to be synchronized with mode-locked lasers, the Pockels cell is driven at MHz rates, with a controllable phase offset relative to excitation laser pulses. The Pockels cell provides a voltage-controlled polarization rotation as a function of time, which allows two time gates to be defined during the camera exposure using the two output polarizations from a polarizing beam splitter (PBS). Pockels cells utilized in EO-FLIM have large crystal dimensions in order to facilitate wide field of view imaging; here we used the largest Pockels cell for EO-FLIM to date, with 17 mm square aperture and 5 mm thickness. In this implementation, it was driven at 80 MHz, compared to previous implementations gated at kHz\cite{bowman_electro-optic_2019} or 40 MHz rates \cite{bowmanscience, bowman_resonant_2021}. 80 MHz modulation corresponds to lifetime estimation sensitivity shifted towards shorter lifetimes (see Supplement for details) and expands the compatibility of the technique with standardly-utilized fluorescence excitation lasers. The Pockels cell module was incorporated into the imaging setup shown schematically in Fig. \ref{fig:setup}a. The fluorescence from the sample was
polarized with a linear polarizer (P1), modulated with the Pockels cell (PC), and separated into gated (G) and ungated (U) images on scientific CMOS camera using a Wollaston prism as a polarizing beam splitter (PBS). The Wollaston prism provided in-line image splitting with a single optical component and may be used provided that it is located near an image plane.

We have
combined the Pockels cell and image splitting optics necessary to implement EO-FLIM with
a commercial selective plane illumination \cite{huisken_optical_2004} light sheet microscope (LaVision Ultramicroscope II). The
sample was excited with an 80 MHz picosecond supercontinuum pulsed laser source and fluorescence is captured
using a large field of view stereomicroscope (Olympus MVX10) with a high numerical aperture (NA) objective  (MVPLAPO 2 XC; 2x magnification and 0.5 NA). During volumetric data collection, a piezoelectric stage displaced the sample in the direction orthogonal to the light sheet. Our sCMOS camera sensor was 3200x3200 pixels and was approximately evenly split between the two gated and ungated images, 1500x1500 pixels, as shown in Fig. \ref{fig:setup}b. This system is capable of performing  light sheet
microscopy across fields of view up to 800 $\mu$m with high collection efficiency. Further details about the experimental setup are given in the Supplement. 

An instrument response function (IRF) was measured by sweeping the controllable phase delay between excitation source and Pockels cell using quenched fluorescein dye solution (see Supplement for more details). The fluorescein was quenched by potassium iodide (KI) at pH 10 in water. We achieved a modulation depth of $\sim$50\% across the wide-field image, computed from the IRF and defined as the difference between maximum and minimium values of the gate intensities normalized to unit sum. The IRF captured information about excitation and gating, and fluorescence decays can be simulated as the convolution between the IRF and an exponential lifetime decay. To determine lifetimes from single frames, the phase offset between laser and the Pockels cell drive was set to an optimal phase delay value for lifetime estimation, where the gate ratio varies steeply as a function of lifetime, particularly around 2 ns. Lifetime was measured by converting a single-frame measurement of image intensity ratio (G/U, the ratio of the pixel intensities in the two polarization channels) to a single-exponential
fluorescence lifetime estimate using a lookup table generated by the convolution described above \cite{bowman_electro-optic_2019}, analyzed at the single experimental phase delay. 

The sensitivity of this lifetime measurement was quantified in Fig. \ref{fig:setup}c, which uses the experimental IRF to calculate from error propagation the theoretical lifetime estimation sensitivity in units of shot noise for the chosen phase delay as a function of lifetime (see Supplement). For pixels with tens of thousands of photons, an F-number of 4 suggests lifetime uncertainty is on the order of 10 ps. Fig. \ref{fig:setup}c demonstrates that this gate position choice deterministically maps gate ratio to lifetime with reasonable sensitivity for the range of relevant lifetimes. We chose the phase delay to best suit mid-range lifetimes, like that of green fluorescent protein, meaning lifetime measurements of fast lifetimes (< 1 ns) are not chosen to be as photon efficient as measurements of lifetimes in the ns range (see Fig. \ref{fig:setup}); however, the chosen relative phase from the signal generator can be adjusted easily whenever desired. The spatial dependence of the F-number is given in the Supplement.

Lifetime acquisition and analysis was performed for every image pixel in parallel. We used a spatially dependent lookup table to accommodate the spatially dependent IRF of the Pockels cell (see Supplement). After image stacks were acquired with dozens or hundreds of frames separated by 1 to 5 microns in z direction, the gated and ungated output images were reduced to a lifetime image with an intensity mask overlaid, mapping photon count to transparency, to maintain biological structural information. Finally, the image was 3D rendered using the Napari \cite{napariref} visualization package. 

The \textit{Arabidopsis} seedlings were grown in 0.5 x Murashige-Skoog (MS) medium with low-melting-point agarose (germination and sample mounting details provided in Supplement). Here we imaged wild type Col-0, multi-FP (Arabidopsis Biological Resource Center [ABRC] stock CS16303), and ER-CFP (ABRC stock CS16250) plants. The multi-FP and single-FP constructs encode plant protein fusions to yellow (YFP), green (GFP), red (RFP) and cyan (CFP) fluorescent proteins. The transgenic line expressing the multi-FPs labels the mitochondria (CoXIV-YFP), plasma membrane (Cam53BD-GFP), plastids (RecA-RFP) and nuclei (Gal4-GFP). The transgenic line expressing ER-CFP labels the endoplasmic reticulum \cite{Nelson2007}.
\medskip

\section{Results}

\begin{figure}[htbp!]
\centering\includegraphics[width=\textwidth]{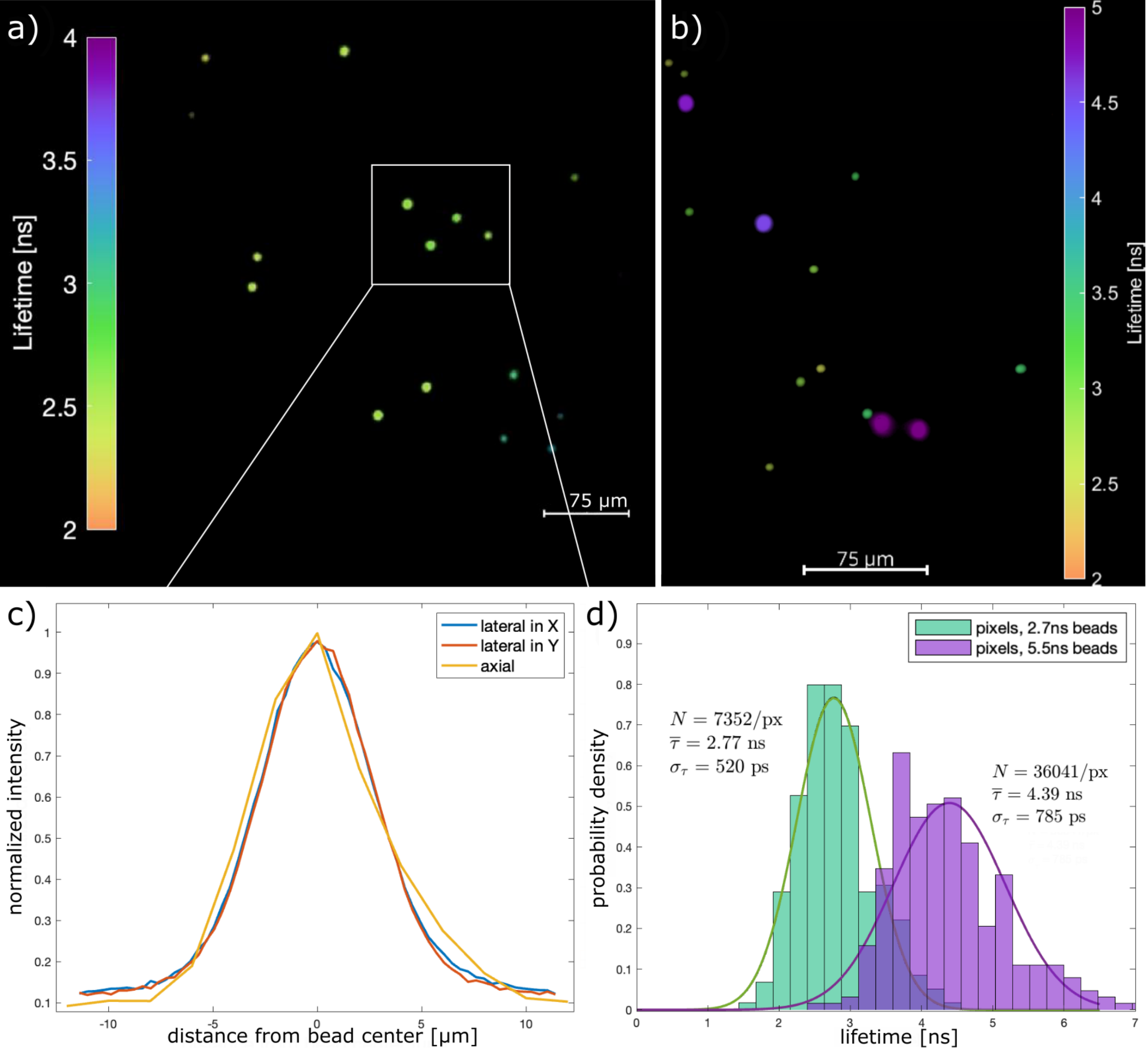}
\caption{Beads used as spatial and lifetime resolution benchmarks. (a) 6.5 micron diameter beads with specified lifetime of 2.7 ns (\href{https://drive.google.com/file/d/1zty6A1FRTEi3X_H7BUqp70GZDlFokRys/view?usp=drive_link}{Visualization 1}). (b) A mixture of 2.7 ns and 5.5 ns beads. The image is masked to show fluorescence intensity information. The mask maps a given photon count to an opacity, so the less bright bead population appears to be smaller when rendered. (c) X, Y, and Z intensity profiles averaged from the central four beads in a), representing the convolution of the beads with X, Y, and Z point spread function. (d) Single-pixel lifetime measurement distribution, normalized as a probability density function, with mean photon counts per pixel, mean lifetime, and lifetime standard deviation.}
\label{fig:beads}
\end{figure}

Fig. \ref{fig:beads} demonstrates volumetric lifetime imaging of a two component mixture of fluorescent beads. We utilized 6.5 $\mu$m diameter commercially available PMMA beads stained with organic dyes from PolyAn GmbH having specified lifetimes of $\sim$2.7 ns and $\sim$5.5 ns \cite{kage2018luminescence}. The beads were suspended in low-melting point agarose. The lifetime was averaged over the entire spatial extent of the bead, and the IRF was calculated locally. The volume in Fig. \ref{fig:beads}a, consisting of 2.7 ns beads, was acquired in 250 frames of 200 millisecond exposure each. The entire volume acquisition took $\sim$
3 minutes. In Fig. \ref{fig:beads}b we show the ability of our system to separate two lifetimes by imaging a mixture of 2.7 and 5.5 ns beads. The 5.5 ns beads are $\sim$2.5x more intense 
than the $\sim$2.7 ns beads \cite{kage2018luminescence}.
The light sheet 1/e$^2$ thickness in our system was measured to be 4 $\mu$m.  In Fig. 2c, we plotted the normalized spatial intensity cross section of imaged beads (averaged over the indicated 4 beads).  The center position of each bead is determined by a circle identification algorithm. The observed lateral (axial) intensity distributions associated with 6.5 $\mu$m diameter beads were 6.9 $\mu$m full width half maximum (7.2 $\mu$m full width half maximum) (Fig. \ref{fig:beads}c). This distribution is the convolution of the point spread function (PSF) and bead size. In Fig. \ref{fig:beads}d, we quantified the precision of single-pixel lifetime measurements, computed from the dataset pictured in Fig. \ref{fig:beads}b.  The purple histogram displays the lifetime values of the individual pixels in the four beads in Fig. \ref{fig:beads}b belonging to the longer-lifetime type, and the green histogram the remaining beads of the second type.

\begin{figure}[htbp]
\centering\includegraphics[width=\textwidth]{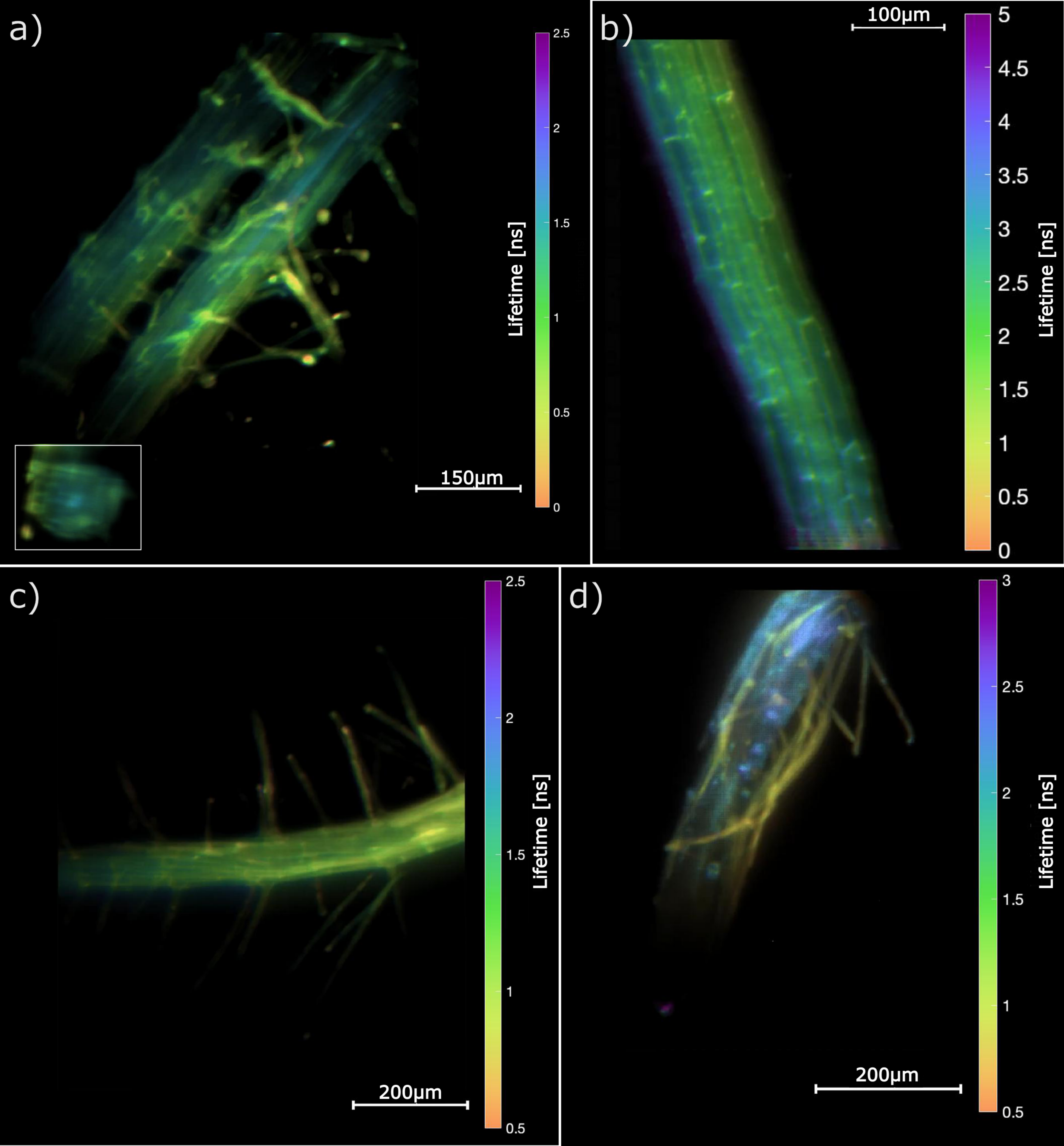}
\caption{\textit{Arabidopsis thaliana} roots labeled with genetically encoded fluorescent protein tags imaged by light sheet EO-FLIM. Volumetric data are represented as a maximum intensity projection along the Z axis. (a-c) Multi-FP plant roots dominantly imaging GFP targeted to the plasma membrane excited at 470/40 nm and emitting at 525/50 nm (\href{https://drive.google.com/file/d/1uJl5Dj5JaLW5IVYcCatv8KtwMsZ1u7_s/view?usp=share_link}{Visualization 2},  \href{https://drive.google.com/file/d/1YScTXwbuZbwt9Nei7oeLoKVzvJyec6uV/view?usp=share_link}{Visualization 3}, \href{https://drive.google.com/file/d/1zrXuhYrlUPX2h_7snz1TxWlEfgCFktob/view?usp=share_link}{Visualization 4}). (a, inset) slice along the axis of the root showing lifetime contrast in the core (\href{https://drive.google.com/file/d/1DDZ9WGLh5iU_uc5bweZaam50jBZxNFSE/view?usp=drive_link}{Visualization 5}). (d) Endoplasmic reticulum labeled with cyan fluorescent protein (CFP) shows contrast between the fluorescent protein structures ($\sim$2.4 ns) and cell wall autofluorescence ($\sim$0.8 ns) (excited 470/40 nm, emitting 525/50 nm) (\href{https://drive.google.com/file/d/1UDCKTMIsH0ir6pAgFz0q8TPHz_h5Qecd/view?usp=share_link}{Visualization 6}).} 
\label{fig:tagged}
\end{figure}

\begin{figure}[htbp]
\centering\includegraphics[width=\textwidth]{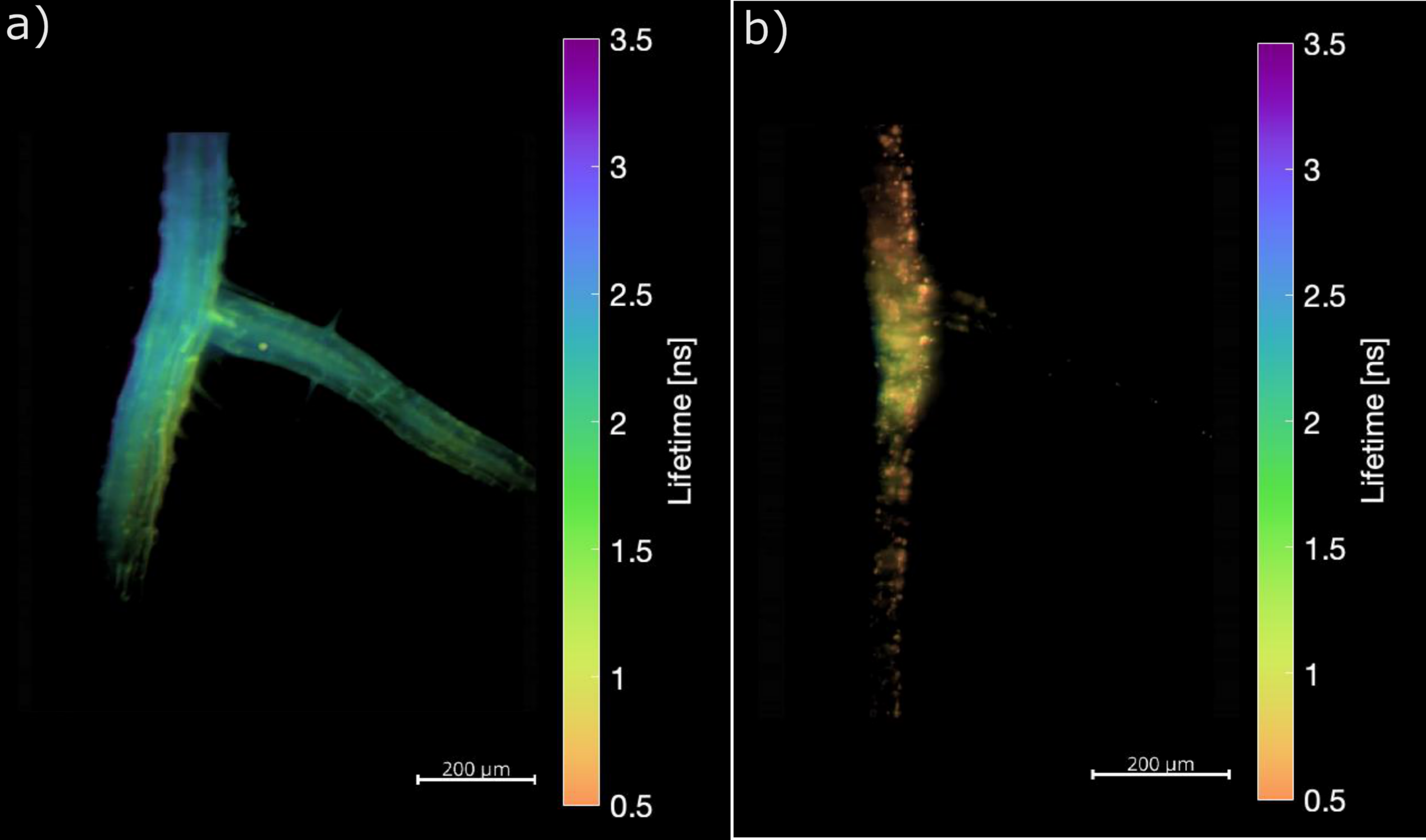}
\caption{Light sheet autofluorescence EO-FLIM captures label-free wild-type seedling root junctions. (a) A junction excited by 470/40 nm emitting at 525/50 nm shows cell wall autofluorescence (\href{https://drive.google.com/file/d/1O-gqGBmTcYGtEsyR7ajGukgNj9iagvkb/view?usp=share_link}{Visualization 7}). (b) A junction excited by 470/40 nm and emitting at 690/50 nm shows internal punctate features which undergo rapid photobleaching (\href{https://drive.google.com/file/d/1opQMm6PtroVYONlq5jZlRkzzVpC8RetF/view?usp=share_link}{Visualization 8}).} 
\label{fig:auto}
\end{figure}

To demonstrate application of LS-EO-FLIM, we imaged \textit{Arabidopsis thaliana} roots with exposure times of 200-500 ms per slice in Fig. \ref{fig:tagged}. Faster acquisitions are also possible depending on sample brightness. Total volume acquisition times ranged from 70 s to 100 s for a full volume scan consisting of 100 acquired planes. In \ref{fig:auto}a, the lifetime-analyzed images with both gates registered to each other are 1816 x 1575 pixels per frame; with a total acquisition time of 313 s for the 300 frames, the lifetime voxel count rate is 2.7$\times 10^6$ photons per second. For the corresponding instrument response function, a total number of 3$\times 10^{12}$ photons (after subtracting background counts) detected in 85 s suggests a 40 GHz photon count rate. The most significant speed limitation is about half a second per image to accommodate the stepping of the sample stage. A comparison of our volume acquisition time to those achieved in other work is shown in the Supplement. 

Volumetric data are represented here as maximum intensity projections, with rotating 3D views linked as Supplementary media. Structures including main root tips, lateral roots, root hairs,  root cells, and  vascular bundles were visualized. We imaged both wild-type roots as well as transgenic roots with fluorescent protein (FP) probes. FP probes enable selective labeling and visualization of distinct sub-cellular compartments with high specificity, which allows lifetime imaging analysis. Autofluorescence, the natural emission of light by biological structures when excited by specific wavelengths, can obscure signals from targeted fluorescent probes and complicate quantitative imaging. However, autofluorescent signals can also provide valuable, noninvasive information about intrinsic cellular components such as cell walls, plastids, and metabolites, offering complementary insights into cellular structure and function in the a native context. As autofluorescence often has much lower fluorescence intensity and spectrally broader emission than FP probes, improving the sensitivity for capturing and interpreting autofluorescent signals in plant tissues is an important application of LS-EO-FLIM.

The dominant signal in our multi-FP plant root images is from plasma membrane localized green fluorescent protein (GFP) due to the large labeled volume in the root, our chosen excitation/emission bands, and limited laser power in the blue spectrum \cite{Kato2008}. Several interesting features were revealed in fluorescence lifetime. For example, we observed a longer lifetime associated with the vascular bundle of the root (Fig. \ref{fig:tagged}a). In Fig. \ref{fig:tagged}d, lifetime provides differentiation between endogenous autofluorescence in wild-type roots and genetically-encoded cyan fluorescent protein targeted to endoplasmic reticulum in transgenic roots, which showed large, longer lifetime masses within the root tissue. This is consistent with the longer expected lifetime of cyan fluorescent protein compared to lignin autofluorescence \cite{Grailhe2006, Borst2010, Dondaldson2013}.

Finally, we noted across many samples an intensity-dependent effect on fluorescence lifetime where the side of the root illuminated more intensely by the light sheet (before it scatters in the root tissue) displayed a shorter lifetime, perhaps due to increased photobleaching. This effect may be more obvious in high-throughput lifetime imaging with LS-EO-FLIM due to the ability to use higher excitation powers without saturating the detector. For our illumination conditions we did not observe structural damage during image acquisition.

We also performed LS-EO-FLIM directly using endogenous autofluorescence of wild-type roots when exciting the sample with blue light (450-490 nm excitation band). Green channel emission dominantly shows the cell walls of wild-type roots which corresponds structurally to the signal observed for transgenic roots labeled at the plasma membrane (Cam53BD-GFP), shown in Fig. \ref{fig:tagged}a. The signal observed in wild type root cell walls is likely due to lignin, which is a structural component of plant cell walls \cite{escamez_fluorescence_2021, donaldson_autofluorescence_2020}. Autofluorescence in a red emission channel shows a concentration of more point-like emitters in the center of the root. Fig. \ref{fig:auto}b reveals internal point-like punctate structures with shorter lifetimes that are particularly sensitive to the incident light intensity and fast to photobleach.

\section{Discussion}
The combination of EO-FLIM with volumetric capture promises to enable applications of label-free microscopy in 3D environments and tissues. For plant imaging, this will provide an improved view of  root dynamics and interactions between roots and microbes in the rhizosphere. More broadly, FLIM readout may be applied to record a variety of fluorescent probes in 3D environments which will enable applications in neuroscience, clinical imaging, and histology.

The demonstrated LS-EO-FLIM capability allows for time-resolved volumetric imaging, where plant volumes can be taken repeatedly at intervals of a few minutes to reveal their dynamics, either internally or in concert with other organisms such as bacteria or fungi. The light sheet modality is optimal for reducing phototoxicity and future improvements will allow long term imaging of plant growth under controlled and natural conditions. The expected low phototoxicity of LS-EO-FLIM is also well suited to two-photon illumination methods that will enable deeper imaging in scattering root tissue. 

The instrument can also be used or adapted for samples with multiple coincident fluorophores at competing signal levels. Even in the scheme presented above, which optimizes speed, the relationship between ratios and the corresponding lifetimes is locally linear, representing an intensity weighted average of lifetime components. To capture multi-exponential information several phase points may be sampled sequentially and analyzed \cite{bowman_resonant_2021}. Such a mode would easily allow multi-exponential volume analysis in <10 minutes of acquisition time, retaining higher photon throughput than alternative acquisition methods.

In summary, we developed an LS-EO-FLIM microscope capable of rendering lifetimes at every pixel of three-dimensional plant volumes in minutes using both endogenous autofluorescence and genetically-encoded probes. This achievement is an extension of recent advances in EO-FLIM that enable wide-field lifetime imaging using resonantly driven Pockels cells to encode nanosecond lifetimes in the ratio of two image intensities. It incorporates the highest achieved resonant drive at 80MHz, corresponding to increased sensitivity at sub-nanosecond lifetimes, and the largest Pockels cell crystals to date, enabling an entirely new regime in field of view. Our system achieves large field-of-view gating on a standard scientific camera, allowing high photon throughput, large dynamic range, and low pixel noise for megapixel FLIM acquisition. Our approach is also compatible with standard 80 MHz mode-locked lasers, making it widely compatible with standard microscopy systems that are in use throughout biological research.

\begin{backmatter}
\bmsection{Funding}
U.S. Department of Energy, Office of Science, Office of Biological and Environmental Research (DE-SC0021976); U.S. National Science Foundation (IOS Grant 1555957, GRFP 1656518).

\bmsection{Acknowledgment} 
We thank Franz Pfanner for assistance with microscope development. We acknowledge Kate Knight for discussions on \textit{Arabidopsis} constructs. 

\bmsection{Disclosures}\\
A.J.B. and M.A.K. are inventors on PCT/US2019/062640, US17/153438, and US17/898093.

\bmsection{Data Availability Statement}
Data underlying the results presented in this paper are not publicly available at this time but may be obtained from the authors upon reasonable request.

\bmsection{Supplemental document}
See Supplement 1 for supporting content.
\end{backmatter}

\label{sec:refs}

\bibliography{lseoflim}






\end{document}


\maketitle

\section{Optical and RF Design}

\begin{figure}[htbp]
\centering
\includegraphics[width=\linewidth]{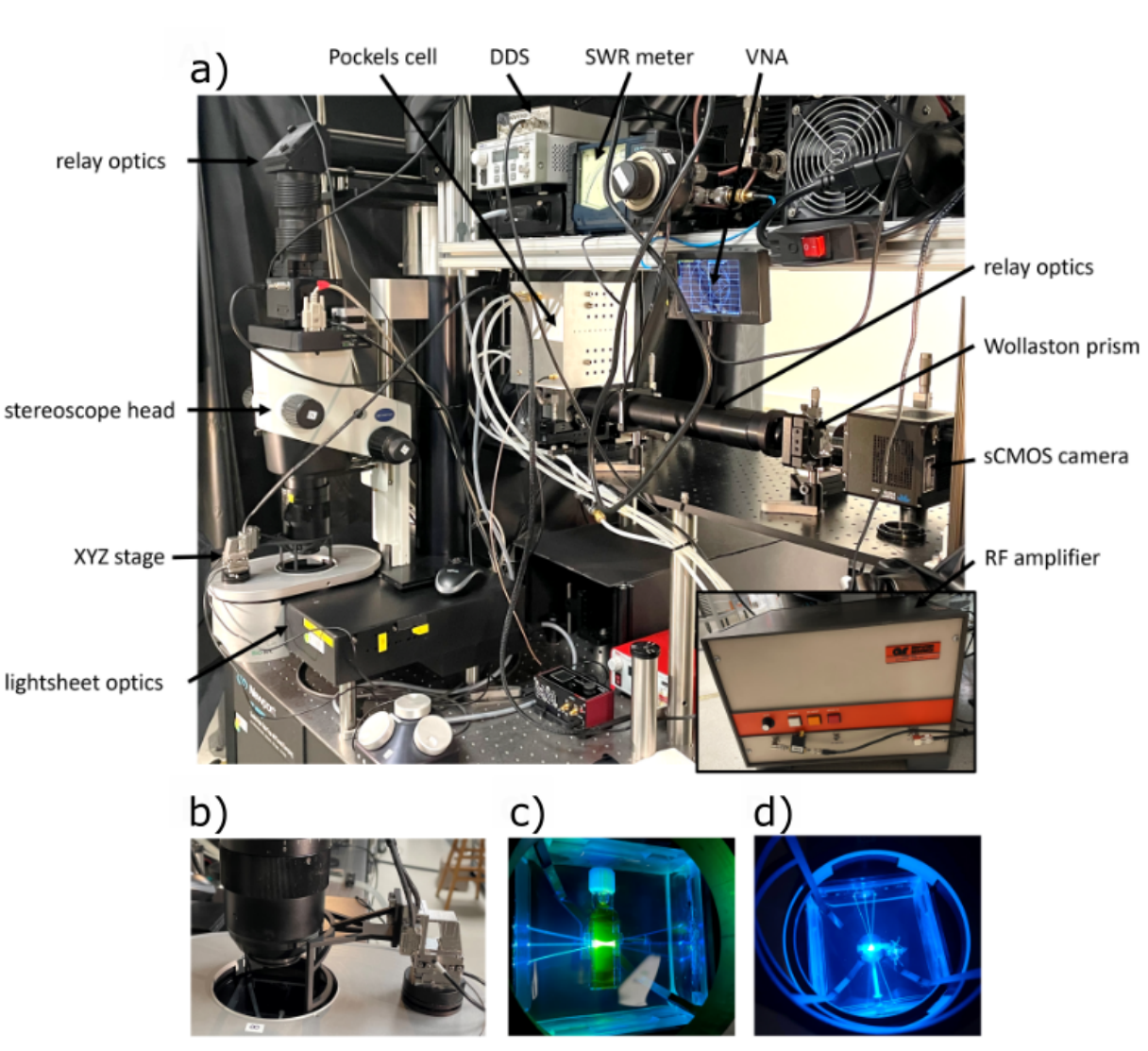}
\caption{(a) Light sheet FLIM setup. The Pockels cell is mounted on XYZ, tip, and tilt stage to
allow positional and axial alignment for optimization of gating performance. The class-A 200
Watt RF power amplifier is shown in the inset. (b) 3D printed sample holder and XYZ sample
stage assembly. (c) A sealed cuvette of quenched fluorescein is used to measure the experimental
instrument response function. (d) \textit{Arabidopsis} samples are grown in cylinders of agarose and
mounted to the 3D printed sample holder. This holder connects to the XYZ stage and allows
motion within the water immersion chamber and Z-scanning for volumetric acquisition.}
\label{fig:setup}
\end{figure}

Our optical system is designed around a large field-of-view stereoscope (Olympus MVX10)
which is part of a commercial selective plane illumination light sheet microscope (LaVision
Ultramicroscope II). The overall setup is pictured in Fig. \ref{fig:setup}A. The optical path was modified to include additional image relays, the Pockels cell, and a Wollaston prism image splitter to implement the EO-FLIM technique. Unpolarized fluorescence is polarized on a linear polarizer, modulated by the Pockels cell, then split on a Wollaston prism in front of an sCMOS camera.

For light sheet microscopy a 2x immersion objective with a water dipping cap is used (MVPLAPO 2 XC; 2x magnification and 0.5NA). The minimum light sheet thickness attainable in this system is 4 $\mu$m. The fluorescence is first polarized on a linear polarizer (LPVISE100-A) and then aligned to be 45 degrees to the fast and slow axes of the Pockels cell
crystals. A dual-crystal lithium tantalate modulator with 17 mm aperture and 5 mm single crystal
thickness (Leysop Ldt.) is employed to enable wide-field imaging while canceling off-axis
birefringence effects \cite{bowmanscience}. This modulator is located near an image plane in the optical relay path. For maximum simplicity, a co-linear design is employed with a Wollaston prism (Karl Lambrecht
Corp. MW2S-20-5) providing in-line polarization splitting onto a large sensor sCMOS camera
(Photometrics Kinetix).

The Pockels cell is incorporated into a resonant transformer tuned to 80 MHz, making it
compatible with standard mode-locked laser sources (Fig. 1). Here we use a supercontinuum laser that provides 100 picosecond pulses (NKT Photonics EXW-12). The laser clock is used as input to a direct digital synthesizer (Novatech 409B) which generates a computer-controlled phase offset. The DDS output is then amplified with a pre-amplifier (MiniCircuits ZHL-1-2W) and a class-A RF power amplifier (Amplifier Research 200L). Power is monitored with an SWR meter (Daiwa CN-901HP).

In our previous EO-FLIM implementations, the Pockels cell introduced zero additional static birefringence. In this implementation, the Pockels cell introduces a $\lambda$/4 phase shift. Therefore the optical gating is at the same frequency as the 80 MHz drive rather than at double the drive frequency as in past resonant EO-FLIM implementations. Using 170 Watts of input RF power, a wide-field modulation depth of 50\% is achieved across the image. The inductor of the resonant transformer is cooled using mineral oil which passes through an air-cooled radiator and gear pump (McMaster-Carr 8220K43). The Pockels cell crystals are also immersed in fluorocarbon coolant (3M Fluorinert FC-40) which is circulated by a compatible chiller (ThermoCube, Solid State Cooling Systems) and the crystal faces and chamber windows are anti-reflection coated to match the refractive index of the coolant.

Care must be taken to ensure adequate time to warm up the tank circuit to avoid thermal drifts in the gating function between IRF and sample measurements. Future designs will improve thermal performance in the tank circuit and achieve higher gating efficiency using lower RF powers. In this system our gating performance at 80 MHz was limited by the choice of PC crystal dimensions and the associated higher $V_\pi$. 

Field of view, depth, and acquisition time are tabulated in Table \ref{tab:shape-functions} for state of the art wide-field light sheet FLIM technologies.

\begin{table}[htbp!]
\caption{Comparison between volumetric FLIM techniques.}
  \label{tab:shape-functions}
  \centering
\resizebox{\textwidth}{!}{\begin{tabular}{l|ccccc}
\hline
Method & Field of view [$\mu m^2$] & Depth [$\mu$m] & Acquisition time \\
\hline
TC-SPC with microchannel plate \cite{hirvonen_lightsheet_2020}& 1500x1500 & 440 &>88 min\\
\hspace{2.8cm}volume sweep \cite{hirvonen_lightsheet_2020} & 1500x1500 & 300&$\sim$5 min \\
modulated CMOS cameras \cite{mitchell_functional_2017} & 280x280 & 100 & < 7 min \\
gated optical intensifiers \cite{funane_selective_2018} & 6000x4000 with stitching 
& 1200 & $\sim$100 min per slice \\
gated optical intensifiers \cite{Greger2011} & 173x132 & 44 &< 1 min\\
TC-SPC with SPAD array \cite{samimi_light-sheet_2023} &175x235 &N/A &$\sim$ 10 s per slice\\
Gated mode SPAD array (static LS)
\cite{Dunsing-Eichenauer2025-cv} & 210x210&80 & 0.18 s per slice \\
\hspace{2.9cm} (scanned LS)
\cite{Dunsing-Eichenauer2025-cv} &  210x210&80 & 1.1 s per slice\\
LS-EO-FLIM (this work) &  800x800&500 &$\sim$ 100 s\\

\end{tabular}}
\end{table}
\pagebreak

\section{\textit{Arabidopsis} sample preparation, mounting, and germination}
Plants are grown in 0.5 x Murashige-Skoog (MS) medium (with Gamborg's vitamins, pH 5.7, Caisson Laboratories MSP06-50LT) with low-melting-point agarose (Thermo Fisher 16520100) inside 5 mm syringes that have had their ends cut off to allow sample extraction before imaging. \textit{Arabidopsis} seeds were sterilized by 70\% ethanol for 1 min, washing three times in sterile water, suspending in 30\% bleach for 30 min, and washing three more times in sterile water. Sterilized seeds are planted into upright syringes, and maintained in a growth chamber (Percival AR-66L) at 50\% humidity, 22 °C, and 100 $\mu$mol/m$^2$/s fluorescent illumination under a 10-h light/14-h dark cycle before imaging experiments inside a plastic tray. Plants are imaged after 1-2 weeks of growth. A wet paper towel is used to provide extra humidity as needed to prevent dehydration of the agarose cylinders. The plant cylinder is extracted from the syringe using the plunger and then adhered to the sample mounting puck using a drop of agarose. Sample mounting pucks and a custom sample holder cage (Fig S1B) were 3D printed from ABS plastic to position the sample under the immersion objective. XYZ sample positioning and light sheet scanning was performed using piezoelectric stages (Thorlabs PD1 for XY axes and PDX1 for Z-axis).

\section{Lifetime estimation and rendering.}
Given the experimentally measured Instrument Response Function (IRF), which accounts for both excitation laser pulse shape and the Pockels cell gating function, a lifetime can be estimated for every pixel of a single frame from the ratio of the two measured image intensities (gated G and ungated U). This estimation is accomplished using a lookup table to convert between the measured $G/U$ intensity ratio $R$ and the corresponding single-exponential lifetime. The lookup table is generated by convolving the measured IRF with the corresponding exponential decay \cite{bowman_electro-optic_2019, bowman_resonant_2021}. 

\begin{figure}[htbp]
\centering
\includegraphics[width=\linewidth]{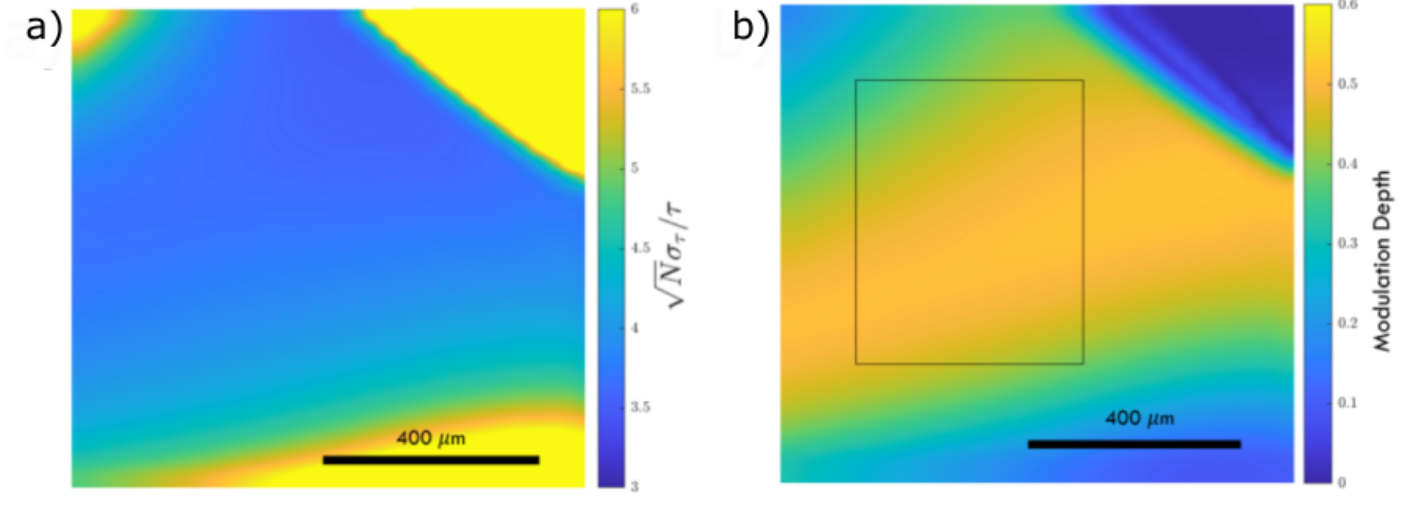}
\caption{(a) Spatial dependence of F-number for 2 ns
lifetime. (b) Spatial dependence of modulation depth across the image. The IRF measured from the indicated region of interest is plotted in the main text Fig. 1.}
\label{fig:irf}
\end{figure}

\begin{figure}[htbp]
\centering
\includegraphics[width=0.68\linewidth]{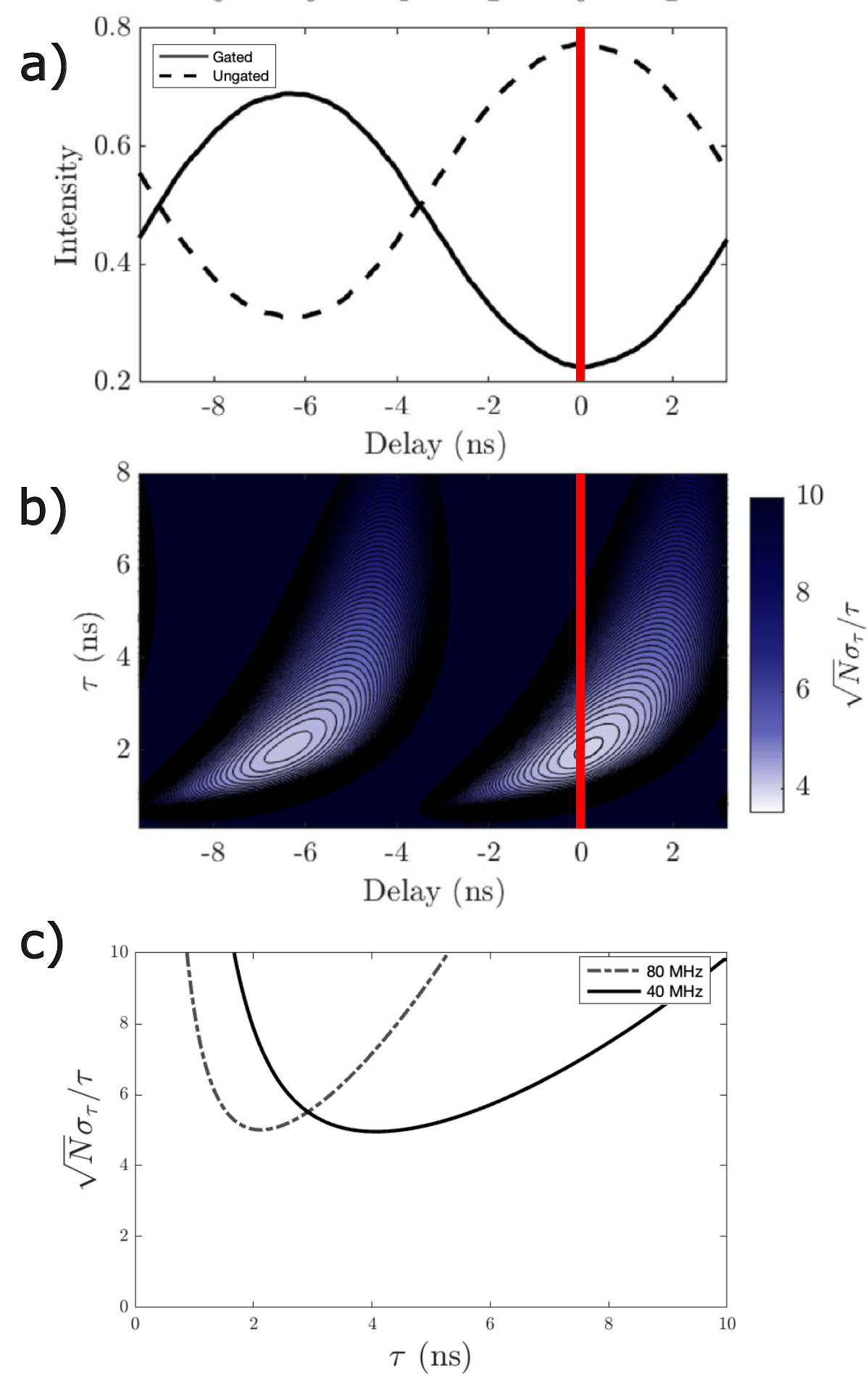}
\caption{(a) Experimental instrument response function (IRF), normalized to unit sum, sweeping delay between Pockels cell and laser and measuring relative intensity for instantaneous fluorescence emitter. During data-taking, delay between Pockels cell and laser fixed at delay indicated by vertical line, set as 0 ns.  (b) Contour plot of F-number, the excess noise factor of the lifetime estimator relative to the shot-noise limit, as a function of delay and lifetime, computed from experimental IRF according to Eq. \ref{eq:errorprop}. (c) F-number at experimental phase delay as a function of lifetime for a waveform with 50\% contrast for each of 40 MHz and 80 MHz Pockels cell modulation, computed from experimental IRF according to Eq. \ref{eq:errorprop}.}
\label{fig:fnumber}
\end{figure}

In order to calibrate the gating function, we image a solution of quenched fluorescein dye
in a vial. A small addition of red fluorescent beads with spectrally-separated fluorescence is
used in order to focus the microscope on the plane of the light sheet during calibration. This
Pockels cell has a spatially varying gating function. In order to correct for this, the image is divided into blocks and a separate lookup
table is calculated for each block in order to convert between measured intensity
ratio and fluorescence lifetime estimate. Supplementary Fig. \ref{fig:irf}a and b plot the spatial dependence of modulation depth and the photon-normalized lifetime estimation sensitivity (F-number) calculated for 2 ns lifetime. For a ratio between the gated and ungated light $R=G/U$, this F-number is computed by error propagation, where

\begin{equation}
F=\frac{\sqrt{N}\sigma_\tau}{\tau}
= \frac{\sqrt{N}}{\tau}\left|\frac{\partial\tau}{\partial R}\right|R\sqrt{\frac{1}{G}+\frac{1}{U}}
\label{eq:errorprop}
\end{equation}

and $F=1$ corresponds to the shot-noise limit. \\
Fig. \ref{fig:fnumber}a presents the experimental IRF with the chosen phase delay indicated. Fig. \ref{fig:fnumber}b shows the F-number as a function of both phase delay and lifetime. Main text Fig. 1b is the slice from Fig. \ref{fig:fnumber}b at this chosen phase delay. Fig. \ref{fig:fnumber}c shows the F-number as a function of lifetime for a 50$\%$ contrast IRF at 40 MHz vs 80 MHz; higher modulation frequency shifts the sensitive region of the instrument to lower lifetimes.\\

The bead images (main text Fig. 2) were generated by computing an average lifetime for each beach using locally estimated IRFs. To analyze plant root data, lookup tables were generated for each of $200\times200$ blocks, meaning each block consisted of $\sim 10\times10$ pixels. Both root and bead data were 3D rendered in Napari after generating lifetime maps and intensity masks for each 2D slice in Matlab. The Napari animation plugin was used to create movies. The Napari-crop plugin was used to slice along any desired axis \cite{napariref}.

\pagebreak

\bibliography{suppl-lseoflim}